\documentclass[a4paper]{jpconf}
\usepackage{graphicx}

\begin{document}

\title{The reactor antineutrino spectrum calculation}

\author{S. V. Silaeva and V. V. Sinev}

\address{Institute for Nuclear Research of Russian Academy of Sciences, Moscow, Russia}

\ead{vsinev@inr.ru}

\begin{abstract}
New fissile isotopes antineutrino spectra ($^{235}$U, $^{238}$U, $^{239}$Pu and $^{241}$Pu) calculation is presented. On base of summation method the toy model was developed. It was shown that total antineutrino number is conserved in framework of given database on individual fragments yields. The analysis of antineutrino spectrum shape says that any presented antineutrino spectrum should satisfy to the total antineutrino number conservation.
\end{abstract}

\section{Introduction}

In 70th and 80th years of the last century there were presented a number of antineutrino spectra calculations [1 $-$ 5], they used limited first databases on fission fragments and were forced to make the models for unknown decay schemes.  

In 80th the measurements of beta-electrons produced by fissile isotopes placed in a reactor core \cite{schrec1} and \cite{schrec2} were presented. Measurements were done for three isotopes that undergo fission by thermal neutrons capturing: $^{235}$U, $^{239}$Pu and $^{241}$Pu. Electron spectra were fitted using 25 hypothetical beta spectra with allowed shape. Then these spectra were transformed to antineutrino ones and they formed antineutrino spectra of fissile isotopes.

Not so far another analysis of the same electron spectra was presented in \cite{muel} and \cite{huber}. They combined fitting of experimental electron spectra and summation method. 

We propose here a new method of reactor antineutrino spectrum calculation ``on-line'' with the simulation of nuclear reactor operating cycle.

During the time passed from previous antineutrino spectra calculation there are appeared new data on decay schemes of the fragments. We used last database on fragments that compiles most known databases \cite{base}. Our computation uses 1039 nuclei with 588 known beta-decayers and 318 unknown.

\section{Summation method}

To get antineutrino spectrum by summation method one needs to summarize all individual antineutrino spectra produced by each beta-transition. Each fragment can have multiple transitions characterized by their branching ratios:  

\begin{equation}
\frac{dN}{dE_{\bar{\nu}}} = \Sigma_n Y_{n}(Z,A)\cdot \Sigma_{n,i}b_{n,i}(^{i}E_{0})P_{\bar{\nu}}(E_{\bar{\nu}},^{i}E_{0},Z),
\end{equation}
where $Y_{n}(Z,A) -$ $n$-th fragment yield per fission, $b_{n,i}(^{i}E_{0}) -$ branching ratio for the transition with $E_{0}$ end point energy, $P_{\bar{\nu}}(E_{\bar{\nu}},^{i}E_{0},Z) -$ is antineutrino spectrum shape for $i$-th branch. Branching ratios are normalized to unity $\Sigma_{i}b_{n,i}(^{i}E_{0}) = 1$. 

Beta spectrum shape $P_{e}(E_{\bar{\nu}},^{i}E_{0},Z)$ is normalized to unity and can be written as
\begin{equation}
P_{e}(E_{e},^{i}E_{0},Z) = K\cdot p_{e}E_{e}\cdot (E_{0}-E_{e})^2\cdot F(Z,E_{e})\cdot C(Z,E_{e})\cdot (1+\delta(Z,A,E_{e}),
\end{equation}
where $K -$ is normalization factor, $p_{e}$ and $E_{e} -$ are momentum and energy of electron, $F(Z,E_{e}) -$ Fermi function accounting Coulomb field of the daughter nucleus, $C(Z,E_{e}) -$ the factor accounting momentum dependence of nucleus matrix element and $\delta(Z,A,E_{e}) -$ is correction factor to the spectrum shape.

The antineutrino spectrum shape $P_{\bar{\nu}}(E_{\bar{\nu}},^{i}E_{0},Z)$ also normalized to unity can be expressed the same way as beta spectrum shape by installing $E_0 - E_{\bar{\nu}}$ on the place of $E_e$. 

Integral of the spectrum (1) gives total antineutrino number per fission 

\begin{equation}
N_{\bar{\nu}} = \int_0^{E_{max}} \frac{dN}{dE_{\bar{\nu}}}(E_{\bar{\nu}})dE_{\bar{\nu}}
\end{equation}
and convolution of the spectrum (1) with inverse beta decay reaction cross section gives total inverse beta decay cross section for a given spectrum

\begin{equation}
\sigma_{tot} = \int_0^{E_{max}} \frac{dN}{dE_{\bar{\nu}}}(E_{\bar{\nu}})\sigma(E_{\bar{\nu}})dE_{\bar{\nu}},
\end{equation}
where $\sigma(E_{\bar{\nu}}) -$ inverse beta decay reaction cross section for antineutrino energy $E_{\bar{\nu}}$ \cite{vogstr}.

Modern point of view on beta decay looks like this: electron and antineutrino are borned in the same moment in the nucleus and all corrections should be applied to the electron and the same way (mirrored) to antineutrino. But there may be another regard on the beta decay process: particles are appeared inside the nucleus with shared energy $Q_{\beta}$, then when leaving the nucleus only the energy of electron is undergo all corrections due to its electric charge and magnetic moment. Energy of antineutrino is not corrected. In this case when calculating $P_{\bar{\nu}}(E_{\bar{\nu}},^{i}E_{0},Z)$ one should omit $F(Z,E_{e})$, $C(Z,E_{e})$ and $\delta(Z,A,E_{e})$ terms that do not affect antineutrino spectrum.   

In the Appendix A we present antineutrino spectra for all reactor fissile isotopes $^{235}$U, $^{238}$U, $^{239}$Pu and $^{241}$Pu. In the Appendix B the cross sections corresponding to the spectra for all reactor fissile isotopes are presented. It is possible to compare our cross section result for $^{235}$U with \cite{giunti} where mean value for all reactor experiments appeared $6.33 \pm 0.08\cdot 10^{-43}$ cm$^2$/fission. Our value is $6.28 \cdot 10^{-43}$ cm$^2$/fission.

\section{Toy-model for reactor antineutrino spectrum}

Using (1), (2) and database \cite{base} the computer code was developed. It allows to compute antineutrino spectrum at any moment of reactor fuel cycle according to the starting fuel composition that can be chosen depending on the goal of computation.

\begin{figure}
\begin{center}
\includegraphics[width=12cm]{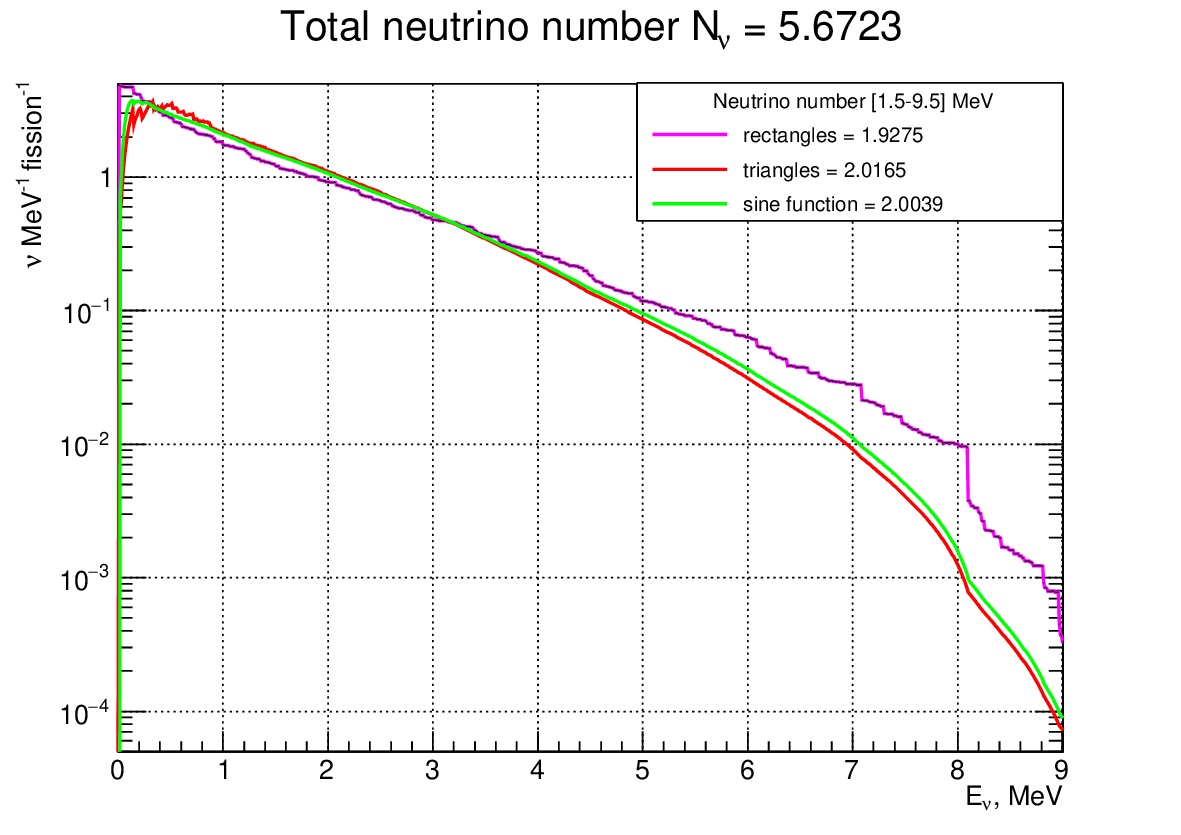}
\end{center}
\caption{\label{figone} Reactor antineutrino spectra in toy-model. Total neutrino number for 1 y irradiation time is 5.67 per fission. Each spectrum has the same integral equals to this value. On the legend panel one also can see antineutrino numbers belonging to energy region 1.5 $-$ 9.5 MeV.}
\end{figure}

Firstly we tested how the spectrum depends on the individual fragment's spectrum shape. There were taken shapes as rectangles, triangles and sine function curve. Results are shown at figure \ref{figone}.  

The computation was done for one year of irradiation time. All spectra are crossing in the point about 3 MeV. This figure shows that if total neutrino number is the same for calculated spectra they should be crossed in some point.

\begin{figure}
\begin{center}
\includegraphics[width=12cm]{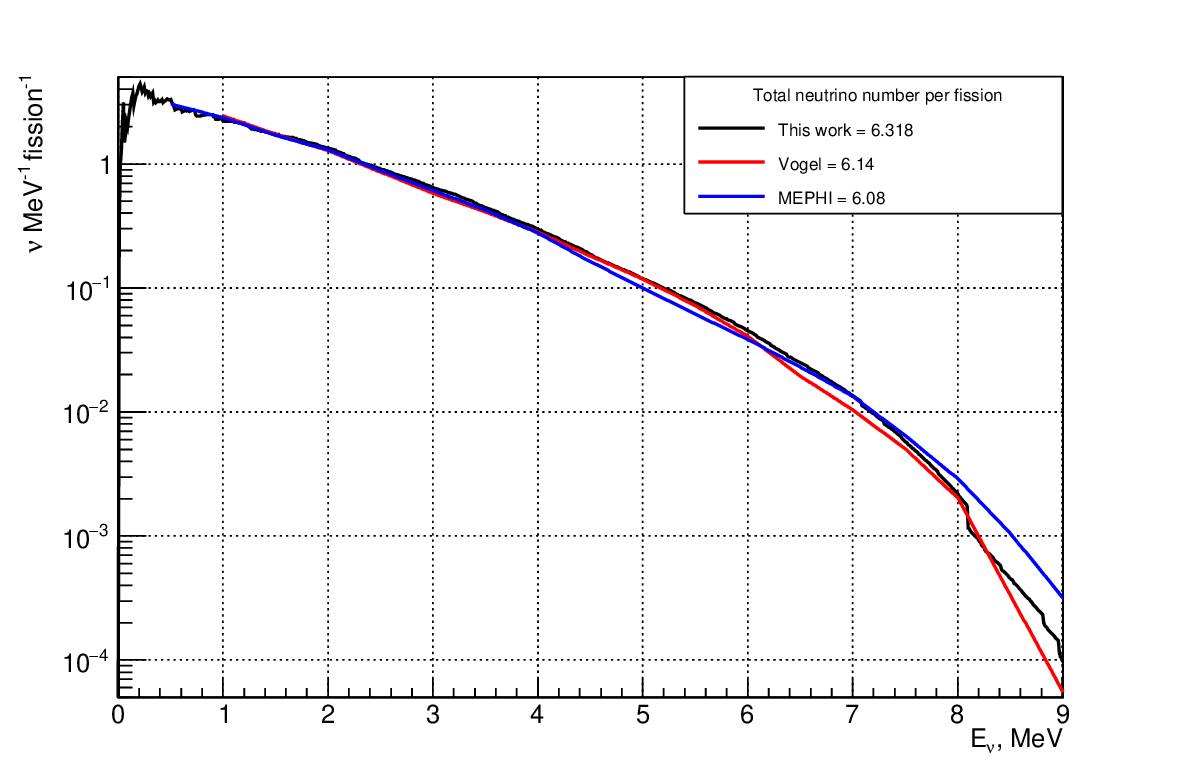}
\end{center}
\caption{\label{figtwo} Antineutrino energy spectra for $^{235}$U taken from \cite{vogel} and \cite{mephi} and the one from this work at infinity exposure time. Total antineutrino numbers per fission are shown for each spectrum in insertion.}
\end{figure}

The comparison was done for antineutrino calculated spectra from \cite{vogel} and \cite{mephi} for infinity exposure time and the spectrum calculated in toy-model for allowed spectrum shape. At figure \ref{figtwo} we see a good agreement for the spectra shape because total antineutrino numbers for all spectra are close to each other (6.14, 6.08 and 6.32).  

Now we compare calculated spectra with the spectrum of $^{235}$U obtained by conversion method in \cite{huber} from experimentally measured beta-electrons from fission fragments in \cite{schrec1}. To do this we should recalculate all spectra to the same exposure time. Usually in reactor core the fuel is presented during 3 years, but it is divided on three parts: one part (1/3) is fresh fuel loaded in reloading period, the second one (1/3) rest from the previous cycle in the core and was moved on a new position and third part (1/3) just spent 2 years in the core and left until next reloading. Let's assume mean time of fuel irradiation time is 2 years. Irradiation time is not exactly exposure time, it is time when fuel is affecting the neutron flux and fragments accumulating and decaying in it. Exposure time is the time in which antineutrinos are detected after the fission happened. In first approximation we can regard these times as equivalent ones.

\begin{figure}
\begin{center}
\includegraphics[width=12cm]{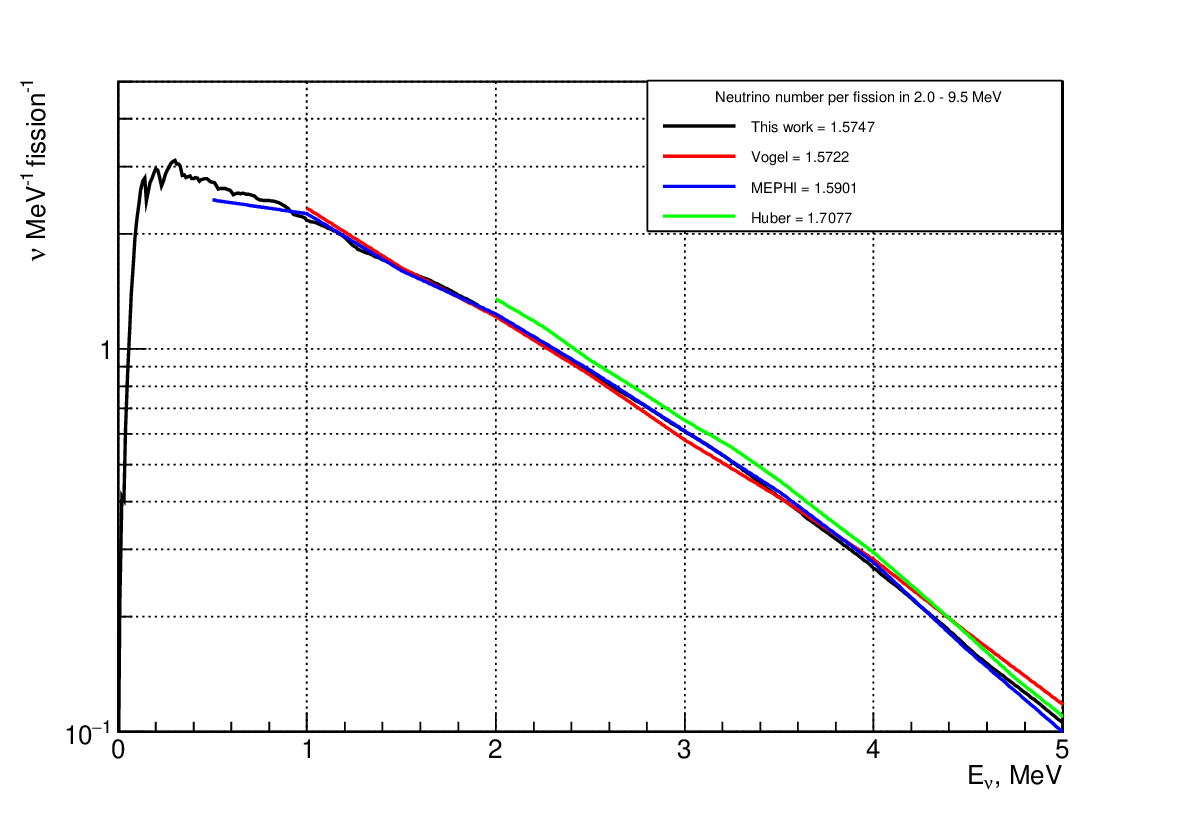}
\end{center}
\caption{\label{figthree} The same antineutrino energy spectra for $^{235}$U as at figure \ref{figtwo} and the spectrum \cite{huber} recalculated to 2 y exposure time. Antineutrino numbers are given for 2$-$9.5 MeV energy region at legend panel. Total antineutrino number for 2 y is 5.68 for calculated spectra only.}
\end{figure}

We have recalculated computed for infinity exposure time spectra in \cite{vogel} and \cite{mephi} to two years of exposure time. The $^{235}$U spectrum from \cite{huber} we also recalculated to two years of irradiation time from 2 days when it was got. The result is shown at figure \ref{figthree}.

The spectrum from \cite{huber} is done starting from 2 MeV only and it is seen that it goes higher than all other spectra in energy region 2$-$4 MeV.

\section{Total antineutrino number per fission}

\begin{figure}
\begin{center}
\includegraphics[width=12cm]{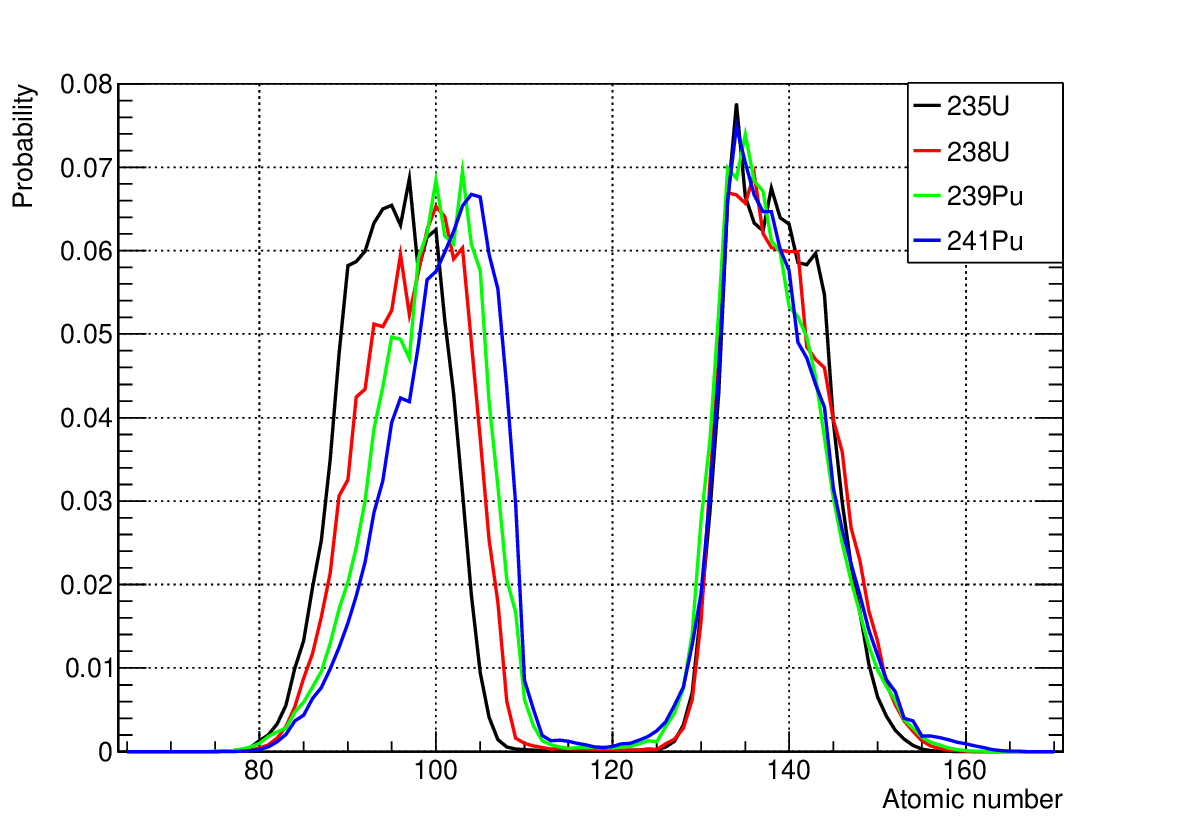}
\end{center}
\caption{\label{figfour} Yields distributions for fission of $^{235}$U, $^{238}$U, $^{239}$Pu and $^{241}$Pu.}
\end{figure}

Integral of antineutrino spectrum gives total antineutrino number for given fissile isotope (1). It depends only on individual fragments yields and does not depend on the fragments spectrum shape. It was shown behigh that all spectra with different individual spectral shape have the same total antineutrino number if were calculated using the same yields database, see figure \ref{figone}. At figure \ref{figfour} one can see cumulative yields distributions for all fissile isotopes. We can note that right peak is practically the same for all fissiles. Possible variations in total antineutrino numbers are due to the left peak which is noticeably different for given fissiles. 

In this section we estimate total neutrino number for the spectrum \cite{huber} under some assumptions because there is now information on the spectrum below 2 MeV. 

As it was mentioned behigh spectra from the same fissile isotopes having the same total neutrinos per fission should be crossed in some point in case of prevailing for high energy part (see figure \ref{figone} and \ref{figtwo}). So, the spectrum \cite{huber} should go down in energy region from 2 down to 1 MeV to reach values of \cite{vogel} one at least to keep similar total antineutrino number. Below 1 MeV spectra should be the same for any author because all fragments are known that formed this part of the spectrum.

\begin{table}[h]
\begin{center}
\caption{\label{tabone} Antineutrinos number $N_{\bar{\nu_e}}$ per fission of $^{235}$U for 2$-$9.5 MeV region and 2 y irradiation time. Total neutrinos per fission 5.687.} 
\vspace{5pt}
\begin{tabular}{|c|c|c|c|}
\hline
 This work & \cite{vogel} & \cite{mephi} & \cite{huber} \\
\hline
  1.5747 & 1.5722 & 1.5901 &1.7077 \\
\hline
\end{tabular}
\end{center}
\end{table}

In this table the result by P. Huber \cite{huber} (1.7077) is 8.15\% higher than the mean value for others (1.579). But it is only a part of total antieutrino spectrum. Total antineutrino number per fission is a physical constant like mean neutron number per fission for fissile isotopes. So, this value could be a criterium of antineutrino spectrum calculation correctness. 

\begin{figure}
\begin{center}
\includegraphics[width=12cm]{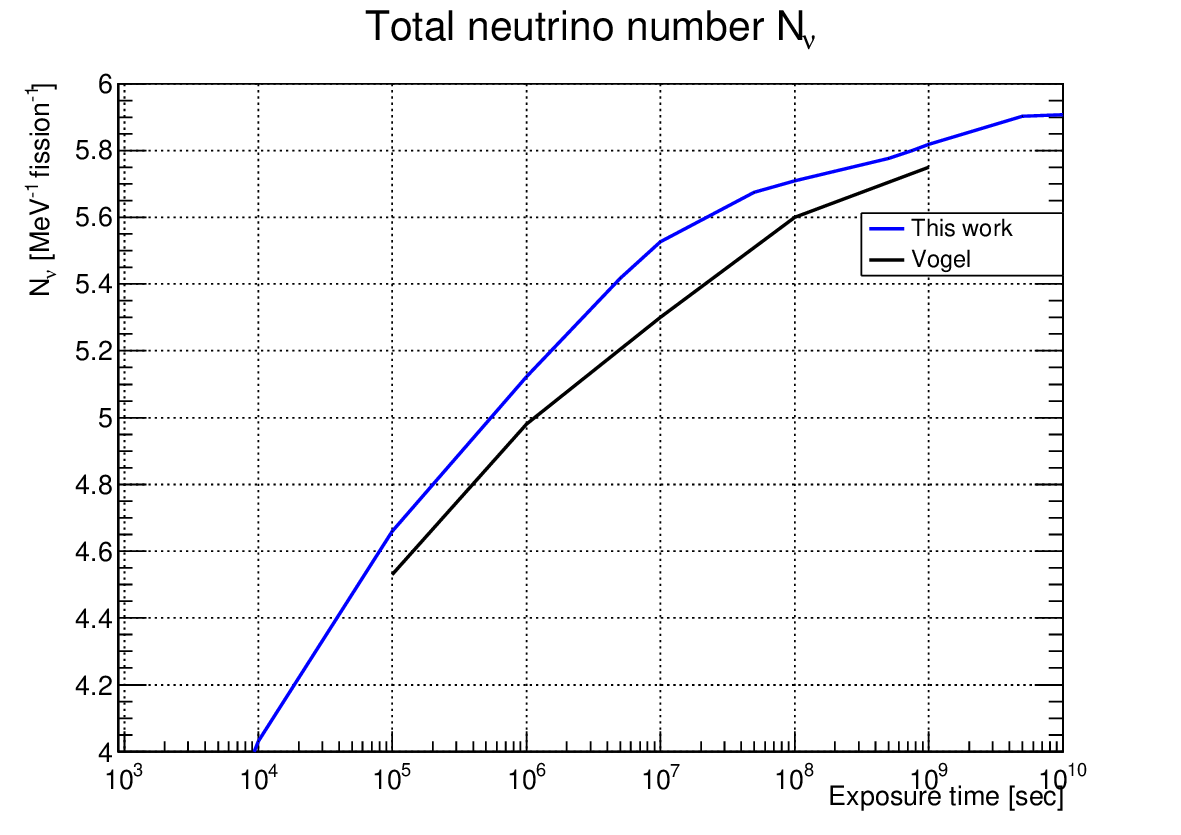}
\end{center}
\caption{\label{figfive} Total neutrino number per fission as a function of exposure time. At infinity time: this work 6.318 and in \cite{vogel} 6.14 antineutrinos per fission of $^{235}$U.}
\end{figure}

At figure \ref{figfive} we show how total antineutrino number per fission changes with exposure time. The data from similar plot in \cite{vogel} are shown here too.

We assume that the spectrum \cite{huber} starts to go down from 2 down to 1 MeV, so we extrapolate it in this region. Than we obtain antineutrino number in wider region 1$-$9.5 MeV and added the soft part of spectra that we assume is the same for all authors because it is formed by known fragments. The similar procedure was done for $^{239}$Pu. The result for $^{235}$U shown in table \ref{tabtwo} and for $^{239}$Pu in table \ref{tabsix}. Last line in these tables shows total antineutrino number for $^{235}$U and $^{239}$Pu that are irradiated 2 years in a core with thermal neutrons. The result for \cite{huber} appeared higher than others. Antineutrino number for \cite{huber} spectrum for $^{235}$U is close to the value for $^{241}$Pu one than to $^{235}$U from other authors what is seen from table \ref{tabthree} where we show total antineutrino numbers for all fissile isotopes. Antineutrino number for $^{239}$Pu is larger than others but not sufficiently, it is still smaller than one for $^{235}$U but close enough.

\begin{table}[h]
\begin{center}
\caption{\label{tabtwo} Antineutrino number $N_{\bar{\nu_e}}$ per fission of $^{235}$U for 2 years of irradiation time.} 
\vspace{5pt}
\begin{tabular}{|c|c|c|c|c|}
\hline
Energy region & This work & \cite{vogel} & \cite{mephi} & \cite{huber} \\
\hline
 1$-$9.5 MeV & 3.2339 & 3.2616 & 3.2532 & 3.5637 \\
 0$-$15 MeV & 5.6875 & 5.527 & 5.473 & 6.0174 \\
\hline
\end{tabular}
\end{center}
\end{table}

\begin{table}[h]
\begin{center}
\caption{\label{tabsix} Antineutrino number $N_{\bar{\nu_e}}$ per fission of $^{239}$Pu for 2 years of irradiation time.} 
\vspace{5pt}
\begin{tabular}{|c|c|c|c|c|}
\hline
Energy region & This work & \cite{vogel} & \cite{mephi} & \cite{huber} \\
\hline
 1$-$9.5 MeV & 2.7439 & 2.7492 & 2.7748 & 2.9779 \\
 0$-$15 MeV & 5.1526 & 5.007 & 4.980 & 5.2357 \\
\hline
\end{tabular}
\end{center}
\end{table}

\begin{table}[h]
\begin{center}
\caption{\label{tabthree} Total antineutrino number $N_{\bar{\nu_e}}$ per fission for 2 years of irradiation time (for $\infty$ time).} 
\vspace{5pt}
\begin{tabular}{|c|c|c|c|}
\hline
Isotope & This work & \cite{vogel} & \cite{mephi} \\
\hline
 $^{235}$U & 5.687 (6.318) & 5.527 (6.14) & 5.473 (6.08) \\
 $^{238}$U & 6.838 (7.416) & 6.528 (7.08) & 6.537 (7.09) \\
 $^{239}$Pu & 5.153 (5.743) & 5.007 (5.58) & 4.980 (5.55) \\
 $^{241}$Pu & 5.967 (6.542) & 5.856 (6.42) & 5.783 (6.34) \\
\hline
\end{tabular}
\end{center}
\end{table}


\section{Conclusion}

The analysis of spectra from \cite{huber} was done using total antineutrino number per fission what is a physical constant for fissile isotope as well as mean neutron number per fission.

The spectra of $^{235}$U and $^{239}$Pu obtained in \cite{huber} were compared with spectra made by summation method and it was shown that the spectra by \cite{huber} does not satisfy to the criterium of total antineutrino number per fission. Total antineutrino number for $^{235}$U appeared close to $^{241}$Pu and for $^{239}$Pu is almost reach the value for $^{235}$U one.

A new fissile isotopes spectra computation is presented. The last database for fragments decay schemes was used. In total 1039 nuclei were used in calculation: 588 known beta-decayers, 318 unknown ones and 133 stable isotopes.


\vspace{19cm}

\section*{Appendix A}

\begin{table}[h]
\begin{center}
\caption{\label{tabfour} Fissile isotopes antineutrino spectra for 2 years of irradiation time.} 
\vspace{5pt}
\begin{tabular}{|c|c|c|c|c|}
\hline
$E_{\nu}$, MeV & $^{235}$U & $^{238}$U & $^{239}$Pu & $^{241}$Pu \\
\hline
 1.00 & 2.196     & 2.372     & 2.025     & 2.292  \\
 1.25 & 1.881     & 2.060     & 1.658     & 1.894  \\
 1.50 & 1.638     & 1.878     & 1.471     & 1.719  \\
 1.75 & 1.442     & 1.697     & 1.296     & 1.545  \\
 2.00 & 1.239     & 1.498     & 1.101     & 1.340  \\
 2.25 & 1.041     & 1.306     & 9.190e-01 & 1.144  \\
 2.50 & 8.719e-01 & 1.135     & 7.642e-01 & 9.688e-01  \\
 2.75 & 7.338e-01 & 9.880e-01 & 6.357e-01 & 8.170e-01  \\
 3.00 & 6.097e-01 & 8.527e-01 & 5.146e-01 & 6.720e-01  \\
 3.25 & 5.079e-01 & 7.322e-01 & 4.146e-01 & 5.536e-01  \\
 3.50 & 4.112e-01 & 6.126e-01 & 3.209e-01 & 4.403e-01  \\
 3.75 & 3.323e-01 & 5.127e-01 & 2.488e-01 & 3.528e-01  \\
 4.00 & 2.680e-01 & 4.286e-01 & 1.938e-01 & 2.826e-01  \\
 4.25 & 2.124e-01 & 3.518e-01 & 1.462e-01 & 2.195e-01  \\
 4.50 & 1.657e-01 & 2.843e-01 & 1.079e-01 & 1.667e-01  \\
 4.75 & 1.328e-01 & 2.325e-01 & 8.193e-02 & 1.286e-01  \\
 5.00 & 1.058e-01 & 1.920e-01 & 6.374e-02 & 1.019e-01  \\
 5.25 & 8.405e-02 & 1.574e-01 & 4.948e-02 & 8.013e-02  \\
 5.50 & 6.591e-02 & 1.275e-01 & 3.781e-02 & 6.190e-02  \\
 5.75 & 5.077e-02 & 1.018e-01 & 2.840e-02 & 4.692e-02  \\
 6.00 & 3.842e-02 & 7.983e-02 & 2.098e-02 & 3.484e-02  \\
 6.25 & 2.837e-02 & 6.182e-02 & 1.504e-02 & 2.544e-02  \\
 6.50 & 2.117e-02 & 4.820e-02 & 1.099e-02 & 1.887e-02  \\
 6.75 & 1.539e-02 & 3.670e-02 & 7.933e-03 & 1.373e-02  \\
 7.00 & 1.083e-02 & 2.697e-02 & 5.402e-03 & 9.448e-03  \\
 7.25 & 7.108e-03 & 1.872e-02 & 3.384e-03 & 5.954e-03  \\
 7.50 & 4.563e-03 & 1.253e-02 & 2.026e-03 & 3.519e-03  \\
 7.75 & 2.793e-03 & 8.110e-03 & 1.159e-03 & 2.017e-03  \\
 8.00 & 1.459e-03 & 4.994e-03 & 5.936e-04 & 1.109e-03  \\
 8.25 & 6.003e-04 & 2.829e-03 & 2.437e-04 & 5.371e-04  \\
 8.50 & 3.448e-04 & 1.696e-03 & 1.310e-04 & 2.783e-04  \\
 8.75 & 1.834e-04 & 1.012e-03 & 5.870e-05 & 1.330e-04  \\
 9.00 & 7.698e-05 & 5.417e-04 & 1.685e-05 & 4.569e-05  \\
 9.25 & 4.550e-05 & 3.359e-04 & 9.914e-06 & 2.649e-05  \\
 9.50 & 2.818e-05 & 2.154e-04 & 6.010e-06 & 1.553e-05  \\
 9.75 & 1.855e-05 & 1.488e-04 & 3.968e-06 & 1.022e-05  \\
10.00 & 1.154e-05 & 9.981e-05 & 2.496e-06 & 6.444e-06  \\
\hline
\end{tabular}
\end{center}
\end{table}

\vspace{5cm}

\section*{Appendix B}

\begin{table}[h]
\begin{center}
\caption{\label{tabfive} Fissile isotopes cross sections $\sigma_f$ in 10$^{-43}$ cm$^2$ for 2 years of irradiation time.} 
\vspace{5pt}
\begin{tabular}{|c|c|c|c|}
\hline
 $^{235}$U & $^{238}$U & $^{239}$Pu & $^{241}$Pu \\
\hline
 6.281  & 10.673 & 4.393 & 6.385  \\
\hline
\end{tabular}
\end{center}
\end{table}

\end{document}